\documentclass[11pt,letterpaper]{article}
\usepackage{systeme} 
\usepackage[utf8]{inputenc}
\usepackage{multirow} 
\usepackage{booktabs} 
\usepackage[english]{babel}
\usepackage{amsmath}
\usepackage{amsfonts}
\usepackage{amssymb}
\usepackage{graphicx}
\usepackage[small,bf]{caption} 
\usepackage[left=3cm,right=3cm,top=2cm,bottom=3cm]{geometry}
\usepackage{framed} 
\usepackage{color} 
\usepackage{wrapfig}\definecolor{shadecolor}{RGB}{220,220,220} 

\author{Jorge Pinochet}
\title{\textbf{Hawking for beginners}}
\begin{document}

\author{Jorge Pinochet$^{*}$\\ \\
 \small{$^{*}$\textit{Facultad de Ciencias Básicas, Departamento de Física. }}\\
   \small{\textit{Universidad Metropolitana de Ciencias de la Educación,}}\\
 \small{\textit{Av. José Pedro Alessandri 774, Ñuñoa, Santiago, Chile.}}\\
 \small{e-mail: jorge.pinochet@umce.cl}\\}

\date{}
\maketitle

\begin{center}\rule{0.9\textwidth}{0.1mm} \end{center}
\begin{abstract}
\noindent Stephen Hawking's most important scientific contribution was his theoretical discovery that “black holes ain’t so black”, since they emit thermal radiation as if they were hot bodies with an absolute temperature known as the Hawking temperature. Although this year marks half a century since Hawking made his discovery, it remains unknown to the vast majority of people. In his popular books, Hawking explains the significance of his discovery in a simple way, using the so-called vacuum quantum fluctuations. The aim of this work is to bring Hawking's discovery to the widest possible audience, especially physics teachers. To do this, we will develop a derivation of the Hawking temperature that fairly faithfully translates Hawking's quantum fluctuation model into the mathematical language used by a typical physics teacher, whether in training or in practice.\\ \\

\noindent \textbf{Keywords}: Black holes, Hawking temperature, quantum vacuum fluctuations, undergraduate students. 

\begin{center}\rule{0.9\textwidth}{0.1mm} \end{center}
\end{abstract}

\maketitle

\section{Introduction}
Stephen Hawking's most important scientific contribution was his theoretical discovery that “black holes ain’t so black” [1], since they emit thermal radiation as if they were hot bodies with an absolute temperature $T_{H}$, known as the \textit{Hawking temperature} [1,2]. This is calculated as:

\begin{equation}
T_{H}=\frac{\hbar c^{3}}{8\pi kGM},
\end{equation}

where $M$ is the mass of the black hole, and $\hbar,c,k,G$ are fundamental physical constants. In his popular books, Hawking set out to explain the meaning and scope of this equation in a simple way; to do this, he resorted to a suggestive physical model that has become very popular, based on a phenomenon known as \textit{quantum vacuum fluctuation} [1,3,4]. Although this year marks half a century since Hawking made his great discovery, he continues to be unknown to the vast majority of people. The objective of this work is to bring Hawking's discovery to the widest possible audience, especially physics teachers, and through them, their students. To this end, a heuristic derivation of $T_{H}$ based on Hawking's quantum fluctuations model is presented. Specifically, we will develop a derivation that, in a semiclassical Newtonian scenario, translates Hawking's model quite faithfully into the mathematical language used by a typical physics teacher, in training or in practice.\\

In the specialized literature it is possible to find more detailed derivations of $T_{H}$ than the one presented here [5–8], as well as others that are mathematically simpler [9–14]. Regarding the former, the main advantage of our derivation is, precisely, its simplicity. Regarding the latter, the most important advantage is that our derivation explicitly incorporates the physical mechanism of Hawking temperature, based on quantum fluctuations of the vacuum, as described by Hawking himself in his popular books.

\section{Quantum fluctuations and the Hawking model}
A black hole is a region of space where gravity is so intense that nothing can escape from its interior, not even light. In the simplest case of a \textit{static black hole}, this spatial region can be intuitively visualised as a sphere, in the centre of which the entire mass of the black hole is compressed (Fig. 1). The surface of this sphere, called the \textit{event horizon} or simply the \textit{horizon}, defines the outer limit of a black hole.\\

Although the horizon has no material existence, it can be imagined as a unidirectional membrane that can only be crossed from the outside to the inside, but never in the opposite direction, even by light. Seen from the outside, the horizon therefore looks completely black. The radius of the horizon or \textit{gravitational radius} is calculated as [5]:

\begin{equation}
R_{g}=\dfrac{2GM}{c^{2}},
\end{equation}

where $M$ is the mass of the black hole, $G=6.67\times 10^{-11}N\cdot m^{2}\cdot kg^{-2}$ is the gravitational constant, and $c\cong 3\times10^{8}m\cdot s^{-1}$ is the speed of light in vacuum.\\

\begin{figure}[h]
  \centering
    \includegraphics[width=0.3\textwidth]{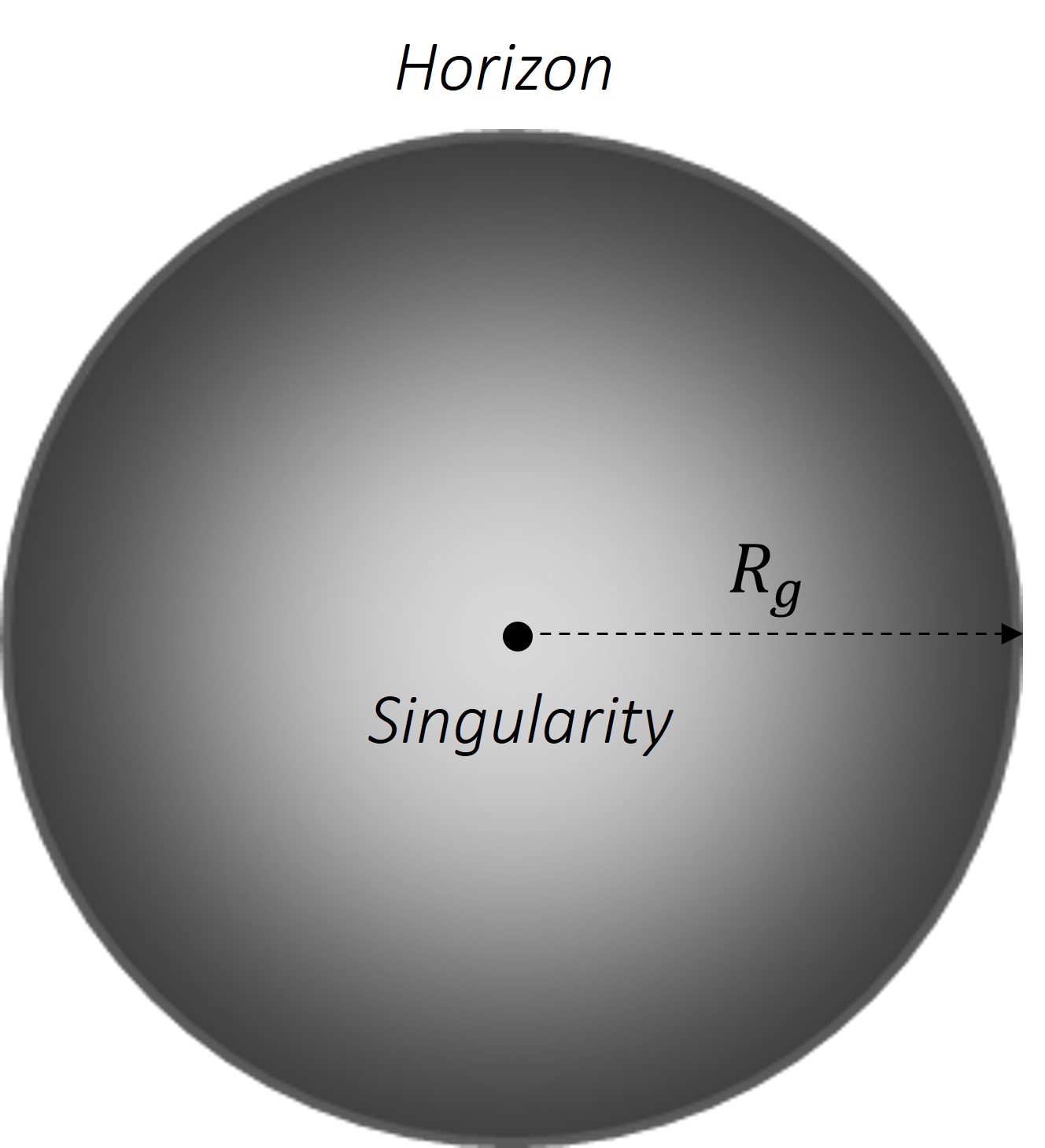}
  \caption{A static black hole is spherically symmetric, has a horizon of radius $R_{g}$ and a central point singularity.}
\end{figure}

What happens if we extend this purely gravitational description of a black hole and incorporate quantum mechanics? This was the question that led Hawking to discover that “black holes ain’t so black.” \\

To delve into this discovery, let's start by introducing the Heisenberg uncertainty principle, one of the most fundamental results of quantum mechanics. According to one of its formulations, if $\Delta E$ is the uncertainty in the energy in a given volume of space, and if $\Delta t$ is the uncertainty in the time during which this energy is measured, the minimum value that the product of these quantities can take is $\Delta E \Delta t=\hbar/2$, where $\hbar=h/2\pi=1.05\times 10^{-34} kg\cdot m^{2}\cdot s^{-1}$ is the reduced Planck constant [6]. \\

The equality $\Delta E \Delta t=\hbar/2$ shows that if the energy present in a vacuum, or its equivalent in mass, were exactly zero at all times and at all points in space, then $\Delta E=0$, which implies that $\hbar=0$. Since this is a contradiction, it is concluded that space cannot remain empty of mass-energy for a finite time. In other words, empty space must fluctuate, continually creating virtual particles that suddenly appear and immediately disappear as they are reabsorbed by the vacuum.\\

Due to the law of conservation of electric charge, virtual particles must arise as particle-antiparticle pairs (remember that for each elementary particle, there is an antiparticle with a charge of equal magnitude but opposite sign), which entails mutual annihilation after a very short time, i.e. below the detection threshold. This ensures that the law of conservation of energy is maintained, since if virtual pairs could be detected, a blatant violation of the law of conservation of energy would occur. The phenomenon in which particle-antiparticle pairs continually appear and disappear is known as quantum vacuum fluctuations, as mentioned in the introduction (see Fig. 2).\\

\begin{figure}[h]
  \centering
    \includegraphics[width=0.3\textwidth]{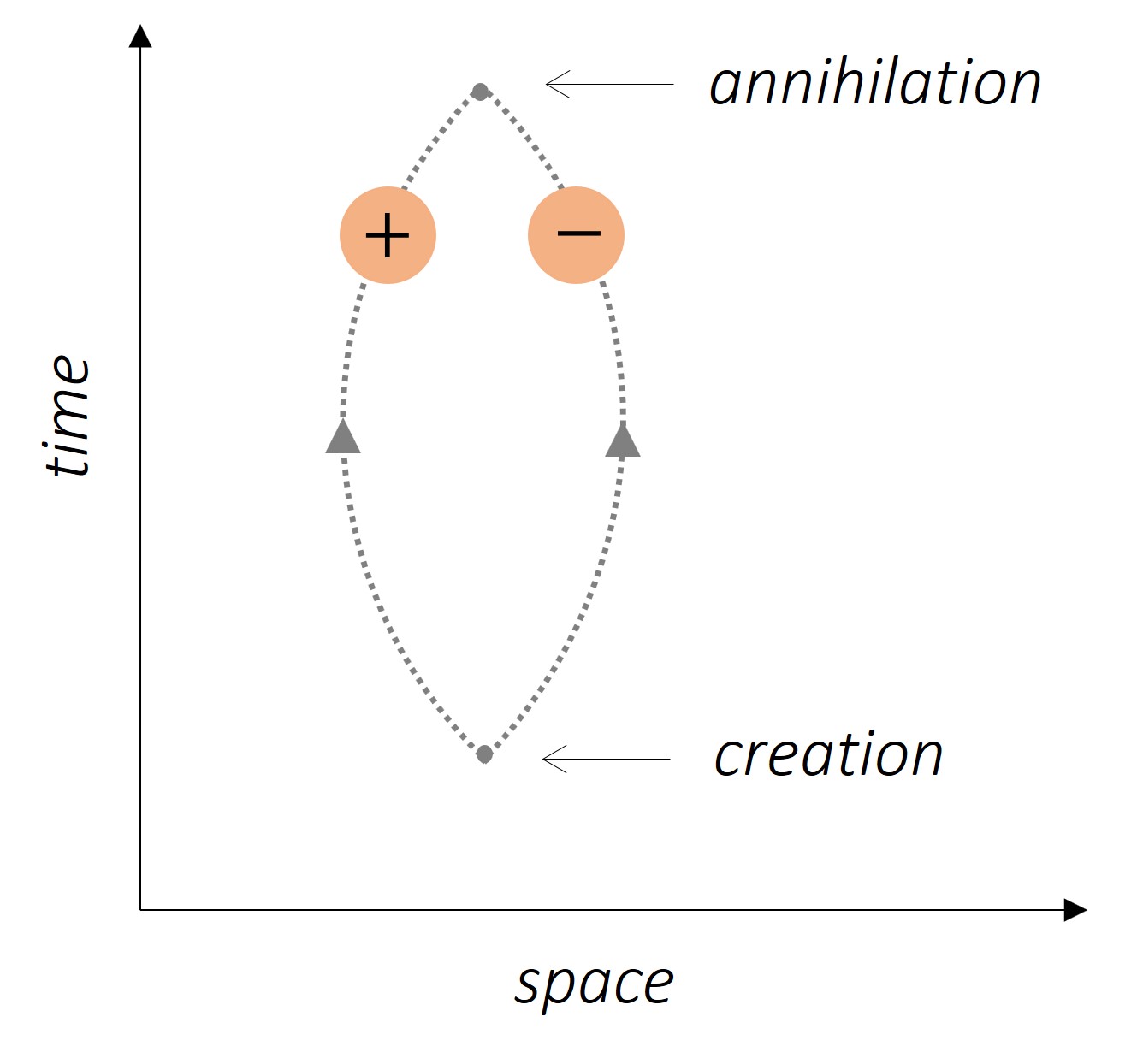}
  \caption{A particle-antiparticle pair suddenly appears in a vacuum. The particles separate and then come back together to annihilate each other and disappear.}
\end{figure}

According to the Heisenberg principle, if $\Delta E=E$ is the uncertainty in the energy of the two virtual particles, the time between the creation and disappearance of the pair is [7]:

\begin{equation}
\Delta t =\frac{\hbar}{2\Delta E}.
\end{equation}

Let us now apply the concept of quantum fluctuation to the description of a black hole. The central idea on which Hawking's model is based is that fluctuations occur continuously at all points in space, and in particular on the outskirts of the event horizon. We focus our attention on one of the myriad virtual pairs that emerge near the horizon, as illustrated in Fig. 3. The intense gravity causes the attractive force on the nearest particle to be greater than on the furthest one; that is, the gravity of the hole produces a \textit{tidal force} that tends to separate the virtual couple. Since this force is very intense near the event horizon, the particles are permanently separated, becoming real particles, because they cannot meet again to annihilate each other. The particle closest to the horizon is absorbed, whereas the furthest one is free to escape to infinity, thanks to the energy injected into it by the tidal force.\\

When repeated over the myriads of quantum fluctuations occurring at all times near the horizon, this process produces a continuous stream of particles emitted randomly in all directions, which a distant observer detects as thermal radiation, or \textit{Hawking radiation}. Hence, for the distant observer, the horizon behaves as a hot body [1,2,4] whose absolute temperature is the Hawking temperature given in Eq. (1).\\

\begin{figure}[h]
  \centering
    \includegraphics[width=0.45\textwidth]{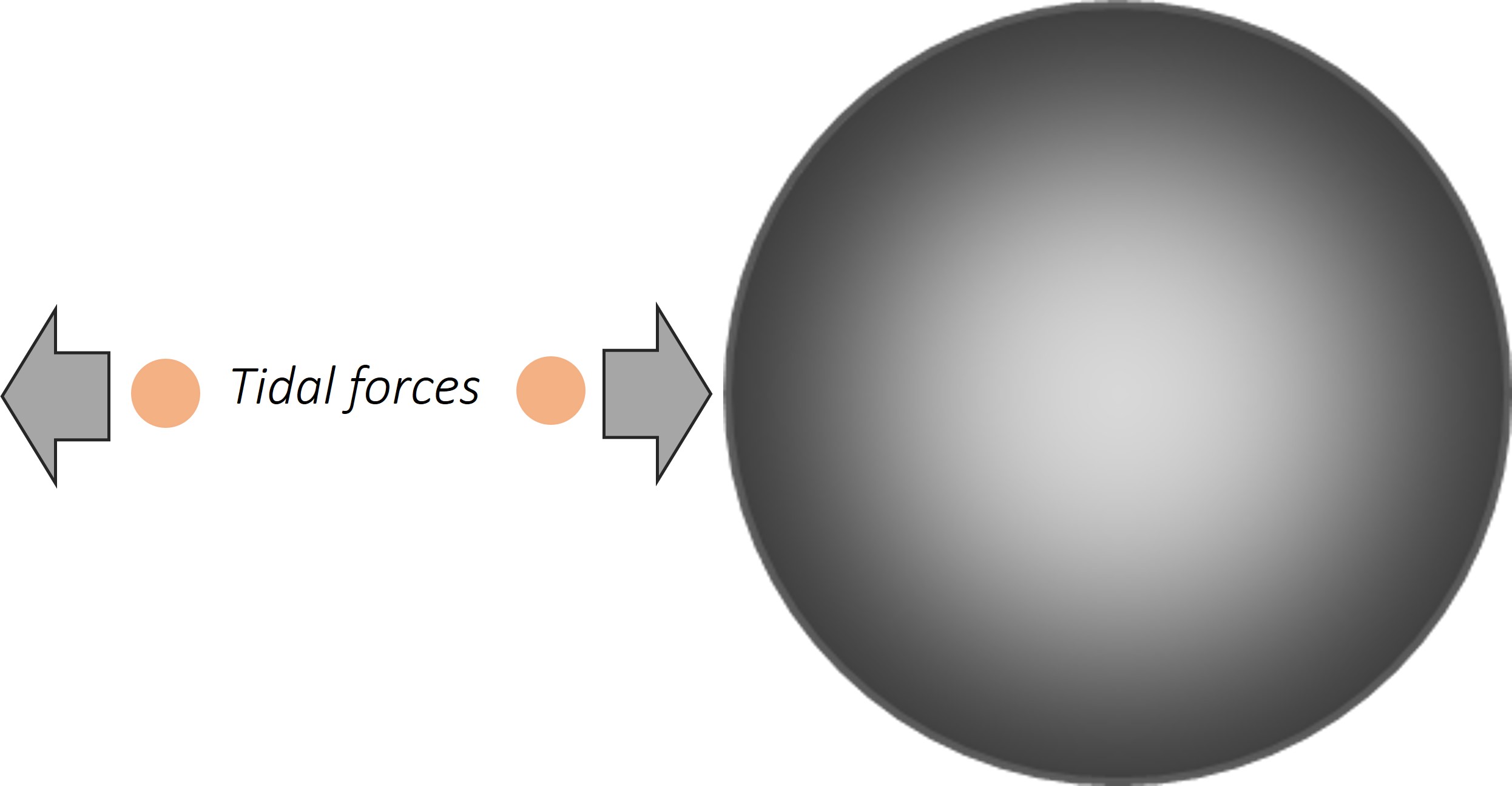}
  \caption{A virtual pair near the horizon is separated by powerful tidal forces. The closest particle is absorbed, but the furthest one is free to escape to infinity.}
\end{figure}

Within the horizon, gravity is so intense that the absorbed particle acquires negative energy $-\varepsilon$. Since energy must be conserved (particles cannot appear out of nowhere), the energy of the emitted particle must be $+\varepsilon$, so $-\varepsilon +\varepsilon=0$. As a result, if the black hole initially has energy $E$ and mass $M=E/c^{2}$, then after emission, its total energy will have been reduced to $E-\varepsilon$, and its mass will therefore have the value $M-\varepsilon/c^{2}$. Then, although this process occurs entirely outside the horizon, it appears to a distant observer as if Hawking radiation were emitted by the black hole itself, at the cost of reducing its mass.\\

Following Hawking's analysis in his popular books, the description we have given of Hawking radiation and temperature applies to massive particles. However, when the particles are massless, the analysis is different. The reader interested in delving deeper into this point will find an accessible discussion in [15] chapter 12.8.

\section{Heuristic derivation of the Hawking temperature}

Since dimensionless numerical factors are unimportant in heuristic calculations, we omit them from the results for simplicity.\\

According to Newtonian physics, the tidal force $F$ exerted by a spherical celestial body of mass $M$ on two particles, each of mass $m$, separated by a distance $l$, is [8,9]: 

\begin{equation}
F = \frac{2GMml}{r^{3}} \sim \frac{GMml}{r^{3}},
\end{equation}

where $r$ is the distance between the particles and the centre of the celestial body. To translate the ideas of the previous section into heuristic calculations, we will use Eq. (4).\\ 

\begin{figure}[h]
  \centering
    \includegraphics[width=0.45\textwidth]{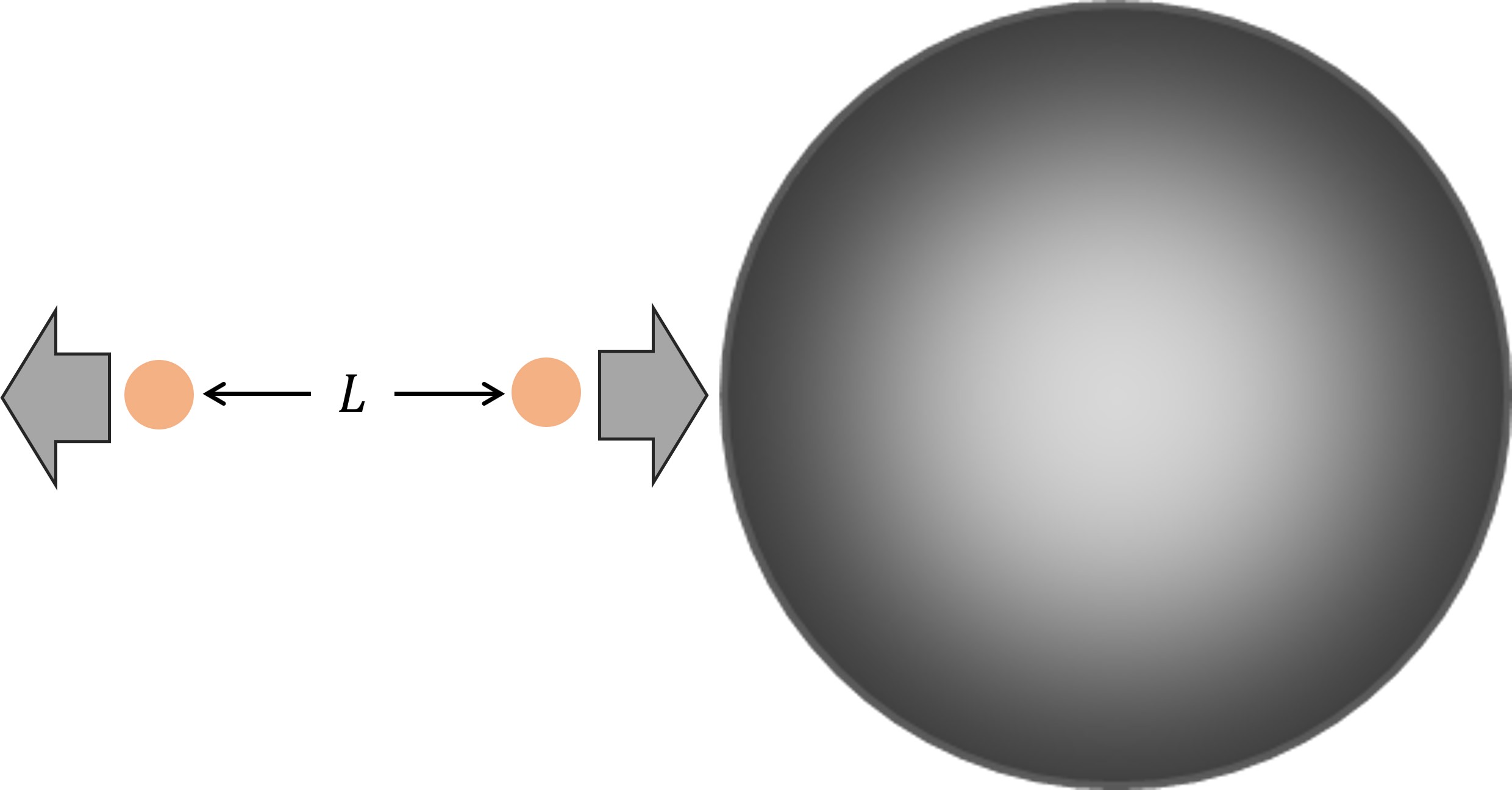}
  \caption{A pair of virtual particles are separated by a distance $L$ due to the action of the tidal force $F$ of the black hole. This force does work $W$.}
\end{figure}

Imagine a black hole of mass $M$ and gravitational radius $R_{g}$. For simplicity, let us consider only the radial direction. Two virtual particles are outside the horizon but very close to it (Figs. 3 and 4). According to the relativistic equivalence between mass and energy, the total energy of the particles is $E=2mc^{2} \sim mc^{2}$. Then, from Eq. (4), the tidal force exerted by the black hole on the pair is:

\begin{equation}
F \sim \frac{GM(E/c^{2})l}{R_{g}^{3}} \sim \frac{c^{4}El}{G^{2}M^{2}},
\end{equation}

where we have taken $r \cong R_{g} \sim GM/c^{2}$, since quantum fluctuations occur very close to the horizon. The work $W$ done by $F$ to separate the particles from an initial distance of zero to a final distance $L$ is: 

\begin{equation}
W \approx \int_{0}^{L} Fdl \sim \int_{0}^{L} \frac{c^{4}El}{G^{2}M^{2}} dl \sim \frac{c^{4}EL^{2}}{G^{2}M^{2}}.
\end{equation}

According to relativistic physics, nothing can move faster than the speed of light in vacuum, $c$. Then, from Eq. (3), we conclude that the maximum distance that can separate a pair of virtual particles is on the order of $c\Delta t \sim c\hbar/E$. Taking $L \sim c\hbar/E$ in Eq. (6), we obtain: 

\begin{equation}
W \sim \frac{c^{4}E}{G^{2}M^{2}}\left( \frac{c\hbar}{E} \right)^{2} = \frac{\hbar^{2}c^{6}}{G^{2}M^{2}E}.
\end{equation}

If virtual particles become real, the energy transferred by $W$ cannot be less than the energy $E$ of the particles themselves. Taking $W \sim E$ in Eq. (7), we obtain $E^{2} \sim \hbar^{2}c^{6}/G^{2}M^{2}$, so that $E \sim \hbar c^{3}/GM$. But Hawking radiation particles are thermal [1,4], and according to statistical mechanics, a thermal particle has an average energy $\langle E \rangle \sim kT$, where $T$ is the absolute temperature, and $k=1.38 \times 10 ^{-23} kg\cdot m^{2}\cdot K^{-1}$ is the Boltzmann constant. Taking $\langle E \rangle = E \sim kT$, we obtain $kT \sim \hbar c^{3}/GM$, and hence $T \sim \hbar c^{3}/kGM$. If we want to convert this proportionality relationship into an equality, we must incorporate the dimensionless constants that would have appeared in an exact calculation. If we group these constants into the dimensionless quantity $\alpha$, we obtain: 

\begin{equation}
T_{H}=\alpha \frac{\hbar c^{3}}{kGM}.
\end{equation}

This result is notable: despite the simplicity of its derivation, it is equal to the Hawking temperature in Eq. (1), which was originally derived by Hawking using much more complex mathematical arguments, allowing him to determine that $\alpha=1/8\pi$ [10,11] (the reader interested in a mathematical derivation whose complexity is halfway between the present work and Hawing's original derivation, can consult [7,8]).\\ 

The smallest (and therefore the hottest) black holes for which there is astronomical evidence are astrophysical black holes, that have masses on the order of the solar mass, $M_{\odot} =1.99 \times10^{30}kg$ (to be precise, due to technical considerations beyond the scope of this paper, the mass $M$ of an astrophysical black hole is $M>3M_{\odot}$). If we take $M \sim M_{\odot}$ in Eq. (8), we obtain $T_{H} \sim 10^{-8} K$, a very low temperature, which is undetectable by astronomical observations [12]. Consequently, it is unlikely that we will be able to empirically verify Hawking's ideas in the short term (for a non-technical discussion of this topic see, for example, [13]). \\ 

Despite this difficulty, the Hawking temperature has been recognised as one of the greatest discoveries in 20th century physics, and represents an important advance in our understanding of the universe. This finding is widely considered to be the main scientific contribution of Hawking, whom we must remember as one of the greatest physicists and science communicator of our time.

\section*{Acknowledgments}
I would like to thank to Daniela Balieiro for their valuable comments in the writing of this paper. 

\section*{References}

[1]	S.W. Hawking, A brief history of time, Bantam Books, New York, 1998.

\vspace{2mm}

[2]	M. C. LoPresto Some Simple Black Hole Thermodynamics: The Physics Teacher: Vol 41, No 5, (n.d.). https://aapt.scitation.org/doi/10.1119/1.1571268 (accessed December 12, 2022).

\vspace{2mm}

[3]	S.W. Hawking, The quantum mechanics of black holes, Scientific American 236 (1976) 34–40.

\vspace{2mm}

[4]	S.W. Hawking, The Universe in a Nutshell, Bantam Books, New York, 2001.

\vspace{2mm}

[5]	A. Martínez-Merino, O. Obregón, M.P. Ryan, Newtonian black holes: Particle production, ”Hawking” temperature, entropies and entropy field equations, (2016).

\vspace{2mm}

[6]	S. Carlip, Black hole thermodynamics, Int. J. Mod. Phys. D 23 (2014) 1430023. https://doi.org/10.1142/S0218271814300237.

\vspace{2mm}

[7]	E.T. Akhmedov, V. Akhmedova, D. Singleton, Hawking temperature in the tunneling picture, Physics Letters B 642 (2006) 124–128. https://doi.org/10.1016/j.physletb.2006.09.028.

\vspace{2mm}

[8]	V. Akhmedova, T. Pilling, A. de Gill, D. Singleton, Temporal contribution to gravitational WKB-like calculations, Physics Letters B 666 (2008) 269–271. https://doi.org/10.1016/j.physletb.2008.07.017.

\vspace{2mm}

[9]	J. Pinochet, The Hawking temperature, the uncertainty principle and quantum black holes, Phys. Educ. 53 (2018) 065004.

\vspace{2mm}

[10] J. Pinochet, “Black holes ain’t so black”: An introduction to the great discoveries of Stephen Hawking, Phys. Educ. 54 (2019) 035014.

\vspace{2mm}

[11] J. Pinochet, Hawking for beginners: a dimensional analysis activity to perform in the classroom, Phys. Educ. 55 (2020) 045018.

\vspace{2mm}

[12] J. Pinochet, Three easy ways to the Hawking temperature, Physics Education 56 (2021) 053001. https://doi.org/10.1088/1361-6552/AC03FC.

\vspace{2mm}

[13] J. Pinochet, Hawking for everyone: commemorating half a century of an unfinished scientific revolution, Phys. Educ. 59 (2024) 055001. https://doi.org/10.1088/1361-6552/ad589c.

\vspace{2mm}

[14] B. Schutz, Gravity from the Ground Up, Cambridge University Press, Cambridge, 2003.

\vspace{2mm}

[15] S.L. Shapiro, S.A. Teukolsky, Black holes, white dwarfs and neutron star. The physics of compact objects, Wiley, Weinheim, 2004.

\end{document}